\documentclass[runningheads]{llncs}

\usepackage[capitalise]{cleveref}
\usepackage{graphicx}
\usepackage{listings}
\usepackage{graphicx}
\usepackage{xspace}
\usepackage{url}
\usepackage{hyphenat}
\usepackage{cite}

\newcommand{\etal}{et al.\xspace}
\newcommand\descitem[1]{\item{\bfseries #1}}
\begin{document}

\title{Quality Assurance of Heterogeneous Applications: The SODALITE Approach}

\titlerunning{The SODALITE Framework for Heterogeneous Applications}

\author{Indika Kumara\inst{1,2} \and Giovanni Quattrocchi\inst{3} \and
Damian Tamburri\inst{1,2} \and
Willem-Jan Van Den Heuvel\inst{1,2} }

\authorrunning{I. Kumara \etal}

\institute{Jheronimus Academy of Data Science (JADS), Netherlands
\and
Eindhoven University of Technology (TUe), Netherlands\\
\and Politecnico di Milano, Italy\\
\email{\{i.p.k.weerasingha.dewage, d.a.tamburri, W.J.A.M.vdnHeuvel\}@tue.nl}
\email{giovanni.quattrocchi@polimi.it}
}

\maketitle 

\begin{abstract}
A key focus of the SODALITE project is to assure the quality and performance of the deployments of applications over heterogeneous Cloud and HPC environments. It offers a set of tools to detect and correct errors, smells, and bugs in the deployment models and their provisioning workflows, and a framework to monitor and refactor deployment model instances at runtime. This paper presents objectives, designs, early results of the quality assurance framework and the refactoring framework.

\end{abstract}

\section{Introduction}
In recent years the global market has seen a tremendous rise in utility computing, which serves as the back-end for practically any new technology, methodology or advancement from healthcare to aerospace. We are entering a new era of heterogeneous, software-defined, high-performance computing environments. In this context, modern distributed applications should be able to utilize heterogeneous Cloud and HPC (High Performance Computing) infrastructures. 

The SODALITE (SOftware Defined AppLication Infrastructures managemenT and Engineering) project aims to support development and operation teams in exploiting heterogeneity. It provides application developers and infrastructure operators with tools that abstract their application and infrastructure requirements to enable simpler and faster development, deployment, operation, and execution of heterogeneous applications.

The SODALITE consortium consists of four academic partners CERTH (Centre for Research and Technology), Jheronimus Academy of Data Science, Polytechnic University of Milan, University of Stuttgart, and five industrial partners ADPT, ATOS, CRAY, XLAB, and IBM. The website and the Github repository of the SODALITE can be found at \textit{sodalite.eu} and \textit{github.com/SODALITE-EU}. The project runs from February 2019 to January 2022.

One of the main objectives of the SODALITE is to assure quality and performance of the deployment models of heterogeneous applications. To this end, the project outcomes includes a taxonomy of errors, bugs, smells, and their resolutions, all pertaining to the deployment and execution of heterogeneous applications (Section 2). Based on this taxonomy, the project builds the tools for verifying and validating deployment models, and predicting smells and bugs in them (Section 3). Once the application is designed and deployed, it is consumed by the end-users. Continuously changing workload and infrastructure resources can make the current deployment suboptimal. Thus, the SODALITE also includes a framework that can monitor the application and its infrastructure, and refactor the deployment as appropriate (Section 4). This paper presents goals, designs, and early results of the aforementioned project outcomes.  
\section{Taxonomy of Smells, Bugs, Errors, and Resolutions}
In the SODALITE, an application deployment is modelled with TOSCA (Topology and Orchestration Specification for Cloud Applications)\cite{R1} and IaC (Infrastructure as Code) \cite{R2, R3} such as Ansible, Chef, and Puppet. A software engineer can inadvertently introduce bugs/smells/errors to the deployment models. A specific result of the SODALITE project is a taxonomy of bugs/smells/errors, and their resolutions for TOSCA and IaC. The taxonomy is to support the development and evaluation of the tools that can predict bugs/smells/errors in heterogeneous application deployments and recommend fixes (see Section 3).

\begin{itemize}
  \descitem{IaC and TOSCA Smells and Resolutions.} We have identified and categorized the smells and their fixes from a multivocal literature review on the best and bad practices for IaC and TOSCA.
  \descitem{IaC and TOSCA Bugs and Resolutions.} A qualitative analysis of commit messages and issue reports is used to derive a taxonomy and data set for IaC bugs and fixes. We are in the latter stages of this study. 
  \descitem{Cloud and HPC Application Bugs/Errors and Resolutions.} We have started a literature review on the bugs and errors pertaining to deployment, operation, and execution of Cloud and HPC applications. 
  \descitem{IaC and TOSCA Errors and Resolutions.} We have identified an initial set of IaC and TOSCA errors and their resolutions from the literature. We will create a complete taxonomy of errors and resolutions based on the results of the above three studies.  
\end{itemize}

\section{The SODALITE Quality Assurance Framework}
\begin{figure*}[!b]
\centering
\includegraphics[width=0.8\textwidth]{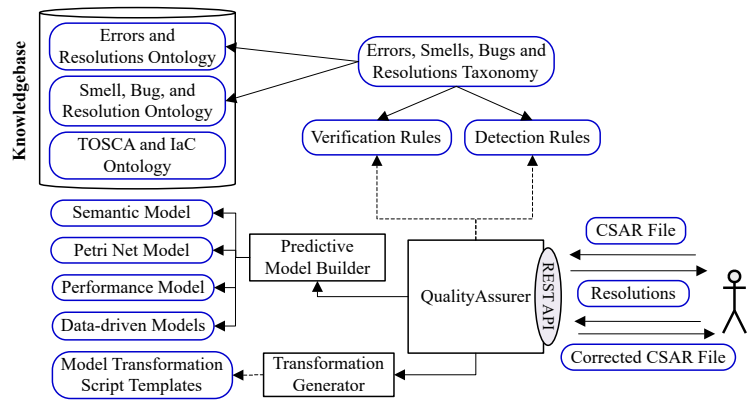}
\caption{The SODALITE Quality Assurance Framework}
\end{figure*}
We provide the developers with a QA framework to find and correct (verification) errors (e.g., inconsistencies), smells, and bugs in a deployment model and its provisioning workflow/plan specified in TOSCA and IaC (the initial focus is on Ansible). The developers can also analyze and validate the performance of an application deployment with our QA framework.

We use the ontological reasoning to verify the constraints over the structures of TOSCA blueprints and IaC scripts. To verify the constraints over the  provisioning workflow (e.g., deadlock detection), we use Petri Net models. To detect smells/bugs, we use three main approaches: informal rules, semantic rules, and data-driven approaches. The informal rules are the detection rules supported by the existing linter tools for IaC (e.g., Ansible-Lint). The semantic rules are reasoning rules over the SODALITE ontologies. The data-driven approaches adopt and further extend the existing machine learning based bug prediction methods developed for general purpose languages. The performance modeling employs a combination of benchmarking/profiling and simulation.

Fig. 1 provides an overview of our QA framework. \textit{Knowledgebase} consists of TOSCA ontology, IaC ontology, errors and resolutions ontology, and smells/bugs and resolutions ontology. We create the first two ontologies based on the TOSCA standard and IaC specifications, and the last two ontologies using the above taxonomies. We use the taxonomies also for defining verification rules and smells/bugs detection rules. \textit{QualityAssurer} takes as input a CSAR (Cloud Service Archive) file consisting of TOSCA, IaC scripts, and performance goals, and uses \textit{Predictive Model Builder} to build the required prediction models. \textit{Predictive Model Builder} can build different types of models: knowledge-based and data-driven models for smells/bugs prediction, knowledge-based models and Petri net models for verification, and statistical models for performance estimation. \textit{QualityAssurer} uses the created models to predict smells/bugs and identify errors and performance violations. It also queries the \textit{Knowledgebase} to recommend the potential fixes. A software engineer can select the desired fixes from the recommendations, and apply the selected fixes to correct the defective artifacts. To ensure consistency and reduce errors, we use model transformations to automate the correction of defective artifacts. Using a template-based approach, \textit{Transformation Generator} generates the required model transformation scripts.    

The early results include the verification of the deployment topologies (TOSCA) using semantic reasoning, and the performance modeling of HPC applications using the data collected from running HPC benchmarks (e.g., LINPACK and STREAM Benchmark) and applying regression analysis on the collected data. The initial support for transforming an Ansible workflow into a Petri Net model has been developed. We extended the Ansible-Lint tool by adding rules for detecting smells (implementation, design, and security) in Ansible. We developed the semantic reasoning for detecting security smells in TOSCA. A curated dataset of Ansible has been created to develop and evaluate data-driven models. We are developing deep learning and NLP based techniques for detecting linguistic smells and module usage issues in IaC. We are extending the CloudSim framework (\textit{cloudbus.org/cloudsim}) for simulating heterogeneous applications.

\section{The SODALITE Refactoring Framework}
\begin{figure*} [!b]
\centering
\includegraphics[width=0.8\textwidth]{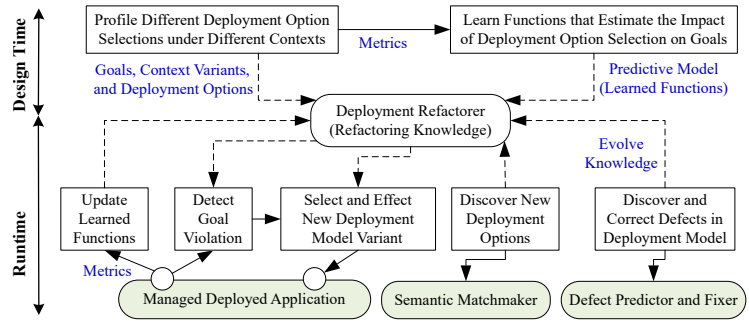}
\caption{The SODALITE Refactoring Framework}
\end{figure*}
The main objective of the predictive deployment refactoring is to refactor or adapt the deployment model of an application at runtime to prevent the violation of the performance goals of the application. The components of an application can be deployed in different ways using heterogeneous resources (e.g., a small VM and a large VM) and deployment patterns (single node, cluster, with or without cache, with or without firewall), resulting alternative deployment options. A valid selection of deployment options results in a valid deployment model variant for the application. The deployment refactoring requires a model that can estimate the impacts of a given deployment option selection on the performance metrics under different contexts such as different workloads. 

Fig. 2 provides an overview of the SODALITE refactoring support. At the design time, we profile deployment variants to collect the data required to build the machine-learning based predictive model. At runtime, the \textit{Refactorer} monitors the deployed application to collect the data and to update the learned model. The predictive model enables the \textit{Refactorer} to predict the potential violations of the application goals, and consequently to find alternative deployment model variants. As the deployment environment evolves, the new resources will be added and the existing resources will be removed or updated. The \textit{Refactorer} discovers new deployment options, the changes to the currently used deployment options, and the bugs introduced by the changes (e.g., performance anti-patterns).

Given the performance goals and the deployment model variant selected at runtime by the \textit{Refactorer}, the SODALITE framework employs distributed control-theoretical planners to further refine the resource allocation of running heterogeneous applications\cite{R4}. For each component deployed in each node of the deployment model, a dedicated \textit{controller} oversees its execution and reallocates CPU and GPU cores without restarting the actual software (i.e., vertical scalability). In addition to the controllers, a \textit{supervisor} is deployed on each node to manage resource contention scenarios that could occur among components running on the same machine. The supervisor governs the allocation of resources according to the actual resources requested (by the controllers), the priority of each component and monitored performance.

We have so far completed the design time part of our framework and implemented the control-theoretical layer. We have developed the methodology to model the deployment variability of heterogeneous applications using the variability modeling techniques. Furthermore, We have developed the approach to profile the different development option selections under different workload ranges, and to use the profiled data to build the machine learning based prediction model. The initial support for semantic matchmaking of deployment options (for discovering new deployment options) also has been developed. The control-theoretical planner can re-configure Kubernetes containers dynamically to maintain response time targets. It currently supports TensorFlow applications that can use both GPUs and CPUs.

\section{Conclusion}
This paper presented an overview of three key tasks of the SODALITE project. During the second year of the project, we plan to complete the taxonomies and the verification support, and to develop data-driven approaches for predicting IaC bugs and for supporting deployment refactoring. During the last year, we plan to complete the rest of the defect prediction tool and the refactorer.
\section*{Acknowledgement}
This paper has been supported by the European Union’s Horizon 2020
research and innovation programme under grant agreement  no. 825480, SODALITE.

\end{document}